\documentclass[aps,prl,groupedaddress,twocolumn]{revtex4-1}

\usepackage{graphicx}
\usepackage{amsfonts}
\usepackage{physics}
\usepackage{color}

\newcommand {\ei} {\textrm{Ei}}

\def\nicefrac#1#2{\leavevmode%
    \raise.5ex\hbox{\small #1}%
    \kern-.1em/\kern-.15em%
    \lower.25ex\hbox{\small #2}}

\begin{document}

\title{Time Operator in Relativistic Quantum Mechanics}

\author{Sina Khorasani}

\affiliation{School of Electrical Engineering, Sharif University of Technology, Tehran, Iran\\
	\'{E}cole Polytechnique F\'{e}d\'{e}rale de Lausanne, Lausanne, CH-1015, Switzerland}
\email{khorasani@sina.sharif.edu; sina.khorasani@epfl.ch}

\begin{abstract}
\textcolor{black}{It is first shown that the Dirac's equation in a relativistic frame could be modified to allow discrete time, in agreement to a recently published upper bound}. Next, an exact self-adjoint $4\times 4$ relativistic time operator for spin-$\frac{1}{2}$ particles is found and the time eigenstates for the non-relativistic case are obtained and discussed. Results confirm the quantum mechanical speculation that particles can indeed occupy negative energy levels with vanishingly small but non-zero probablity, contrary to the general expectation from classical physics. Hence, Wolfgang Pauli's objection regarding the existence of a self-adjoint time operator is fully resolved. It is shown that using the time operator, a bosonic field referred here to as energons may be created, whose number state representations in non-relativistic momentum space can be explicitly found.

\end{abstract}


\maketitle

\section*{Time Spectra}

In their recent paper, Faizal \textit{et al.} \cite{1} suggest that an operator for time could be defined and introduced to the non-relativistic Schr\"{o}dinger's equation, which would recast it in the form
\begin{equation}\label{eq1}
  i \hbar \frac{\partial}{\partial t}  \ket{\psi}+\alpha\hbar^2\frac{\partial^2}{\partial t^2} \ket{\psi}=\mathbb{H} \ket{\psi},
  \end{equation}
\noindent
where $\alpha$ with the dimension of inverse energy is introduced through appropriate deformed algebra. The resulting feature of (\ref{eq1}) is that there will result two distinct energy eigenvalues, and thus the the time-coordinate would acquire a minimum time-step feature. While the subject of time has been quite controversial, as nicely summarized by Muller \cite{2,3,4}, the existence of a Hermitian time-operator has been a subject of long debate.

Pauli \cite{5,6} was the first to point out that it was impossible to have such an operator of time. He did not present a well-founded proof, however, and his reason was that the time needed to be a continuous variable while energy eigenstates must be bounded from below \cite{7}. This viewpoint has
been followed in many of the later studies which have been discussed in several recent reviews of this subject \cite{8,9,10}. However, it has been rigorously shown \cite{11,12} that a self-adjoint Hermitian Time-of-Arrival operator can be indeed accurately defined and used. Also, in his comment \cite{13} and earlier works \cite{14}, Sidharth also points out the possibility of discreteness at the Compton Scale to develop a consistent theoretical framework for cosmology. It is here shown that this is in fact quite possible in the fully relativistic picture of quantum mechanics, without retaining to such a non-standard algebra.

The $4\times4$ Dirac's equation subject to a potential reads
\begin{equation}\label{eq2}
\begin{bmatrix}
 mc^2+[\mathbb{V}]^{+} & c\vec{\sigma}\cdot\hat{\bf p} \\
   c\vec{\sigma}\cdot\hat{\bf p} & -mc^2+[\mathbb{V}]^{-}
 \end{bmatrix}
\begin{Bmatrix}
 \ket{\psi^{+}}\\
 \ket{\psi^{-}}
 \end{Bmatrix}
 =
 i\hbar\frac{\partial}{\partial t}
\begin{Bmatrix}
 \ket{\psi^{+}}\\
 \ket{\psi^{-}}
 \end{Bmatrix},
\end{equation}
\noindent
where $[\mathbb{V}^{\pm}]$ are appropriate perturbing spin-matrix potentials for particles and antiparticles, and $\ket{\psi^{\pm}}$ are spinors. The vector $\vec{\sigma}$ is defined as $\vec{\sigma}=\sigma_x\hat{x}+\sigma_y\hat{y}+\sigma_z\hat{z}$ where $\sigma_j$ with $j=x,y,z$ are $2\times 2$ Pauli's spin matrices \cite{5,7}. The $2\times 2$ identity matrix $[\textrm{I}]=\sigma_0$ has not been shown for the convenience of notation. The above equation may be rearranged by taking time-derivatives from each row of (\ref{eq2}) and straightforward substitution, to yield the modified $4\times4$ Schr\"{o}dinger's equation of the type
\begin{equation}\label{eq3}
   i \hbar \frac{\partial}{\partial t}  \ket{\vec{\psi}}+\alpha\hbar^2\frac{\partial^2}{\partial t^2} \ket{\vec{\psi}}=[\mathbb{H}] \ket{\vec{\psi}},
\end{equation}
\noindent
similar to (\ref{eq1}) with $\alpha=1/mc^2$. Curiously, this value of $\alpha$ satisfies the expected upper bound of $7.2\times 10^{23} {\rm J}^{-1}$  \cite{1} by many orders of magnitude, even for a particle as light as electron for which $\alpha=1.2\times 10^{13} {\rm J}^{-1}$. Also, one would obtain
\begin{eqnarray}\label{eq4}
 \ket{\vec{\psi}}&=&\begin{Bmatrix}
 \ket{\psi^{+}}\\
 \ket{\psi^{-}}
 \end{Bmatrix},
 \\
 \nonumber
 [\mathbb{H}]&=&\begin{bmatrix}
 -[\mathbb{V}^{+}]-\frac{1}{m}\hat{p}^2 & c\vec{\sigma}\cdot\hat{\bf p} \\
   c\vec{\sigma}\cdot\hat{\bf p} & -[\mathbb{V}^{-}]+\frac{1}{m}\hat{p}^2
 \end{bmatrix},
\end{eqnarray}
\noindent
correct to the first order in $[\mathbb{V}^\pm]$. This Hamiltonian will clearly in general lead to four distinct eigenvalues by using the perturbation technique, for particles and anti-particles if $[\mathbb{V}^{+}]\neq -[\mathbb{V}^{-}]$ (broken fundamental CPT symmetry) and also $[\mathbb{V}^{\pm}]$ are non-degenerate themselves. But if only $[\mathbb{V}^{\pm}]$ are kept as non-degenerate (for instance, by having different interactions for different spins) then one could obtain the expected split between the two eigenmodes of the system for either particles or anti-particles. In that case, the concept of time crystals and a minimum time-step as discussed and introduced therein \cite{1} would be immediately plausible.

\section*{Time Operator}

The question of existence of an algebraic form for a self-adjoint time operator of Dirac's equation has been unsolved for a long time, and mostly people have come up with approximate solutions \cite{15,16}. Wang \textit{et al.} \cite{17,18} argued that the correct time operator of Dirac's equation should not be conjugate to the original Hamiltonian, and followed a time of arrival formalism to proceed with a closed form. Uncertainty relationships have been reviewed in a recent survey article \cite{19} and connections of the theory of time to various applications such as tunneling \cite{20} has been discussed. 

Interestingly, it is possible to construct a $4\times 4$ time operator $\mathbb{T}$ for spin-$\frac{1}{2}$ particles in free space, in such a way that the commutator is exactly equal to $i\hbar$, subject to the condition that the angular momentum operator $\hat{\textbf{L}}=\hat{\textbf{r}}\times\hat{\textbf{p}}$ identically vanishes. This latter criterion can be physically satisfied at ease for the case of propagation of particles in free space without presence of electromagnetic fields, while preserving their intrinsic spin property. For the case of a non-vanishing angular momentum, analytical construction of the time operator seems not to be feasible. Even though, this is to the best knowledge of the author, the first conclusive and unambiguous determination of an analytical time operator, which is also compatible with the relativistic Dirac's picture and particle spin. 

This will give rise, after tedious algebra done by hand, to the expression of the relativistic time operator
\begin{equation}\label{eq5}
\mathbb{T}=\begin{bmatrix}
             \frac{m}{6\hat{p}^2}(3\hat{\bf p}\cdot\hat{\bf r}-\hat{\bf r}\cdot\hat{\bf p}) & \frac{1}{3c}\vec{\sigma}\cdot\hat{\bf r} \\
             \frac{1}{3c}\vec{\sigma}\cdot\hat{\bf r} & -\frac{m}{6\hat{p}^2}(3\hat{\bf p}\cdot\hat{\bf r}-\hat{\bf r}\cdot\hat{\bf p})
           \end{bmatrix},
\end{equation}
\noindent
where the $4\times 4$ Dirac Hamiltonian $\mathbb{E}$ is
\begin{equation}\label{eq6}
\mathbb{E}=\begin{bmatrix}
 mc^2 & c\vec{\sigma}\cdot\hat{\bf p} \\
   c\vec{\sigma}\cdot\hat{\bf p} & -mc^2
 \end{bmatrix}.
\end{equation}
\noindent
In derivation of (\ref{eq5}) one should make use of the identities $(\vec{\sigma}\cdot\textbf{A})(\vec{\sigma}\cdot\textbf{B})=\textbf{A}\cdot\textbf{B}+i\vec{\sigma}\cdot(\textbf{A}\times\textbf{B})$, and $\hat{\textbf{r}}\cdot\hat{\textbf{p}}-\hat{\textbf{p}}\cdot\hat{\textbf{r}}=3i\hbar$. Despite the fact that direct derivation of (\ref{eq5}) is very lengthy, it is not difficult to check directly by substitution that together (\ref{eq6}) and vanishing angular momentum they indeed exactly satisfy $[\mathbb{T},\mathbb{E}]=i\hbar$.

The conjugate relationships are now simple given by \cite{6,20a}
\begin{eqnarray}\label{eq7}
   +i \hbar \frac{\partial}{\partial t}  \ket{\vec{\psi}}=[\mathbb{E}] \ket{\vec{\psi}},\\ \nonumber
   -i \hbar \frac{\partial}{\partial e}  \ket{\vec{\chi}}=[\mathbb{T}] \ket{\vec{\chi}}.
\end{eqnarray}
\noindent
Here, $\ket{\vec{\chi}}$ is the $4\times 1$ state ket of the system in energy representation, as opposed to the $\ket{\vec{\psi}}$ in (\ref{eq4}) is the familiar $4\times 1$ state ket of the system in time representation. Hence, the energy and time eigenstates may be found be solution of the equations \cite{6,20a}
\begin{eqnarray}\label{eq8}
E \ket{\vec{\psi}_E}&=[\mathbb{E}] \ket{\vec{\psi}_E},\\ \nonumber
T \ket{\vec{\chi}_T}&=[\mathbb{T}] \ket{\vec{\chi}_T},
\end{eqnarray}
\noindent
where $E$ and $T$ are energy and time eigenvalues, and $\ket{\vec{\psi}(t)}=\textrm{exp}(-\frac{i}{\hbar}Et)\ket{\vec{\psi}_E}$ is the \emph{time-dependent energy eigenstate}, while $\ket{\vec{\chi}(e)}=\textrm{exp}(+\frac{i}{\hbar}Te)\ket{\vec{\chi}_T}$ is the \emph{energy-dependent time-eigenstate}. These are evidently related through the Fourier transformation pairs as \cite{6,20a}
\begin{eqnarray}\label{eq9}
\ket{\vec{\chi}_T}&=\frac{1}{\sqrt{2\pi\hbar}}\int_{-\infty}^{+\infty}e^{-\frac{i}{\hbar}ET}\ket{\vec{\psi}_E}dE,\\ \nonumber
\ket{\vec{\psi}_E}&=\frac{1}{\sqrt{2\pi\hbar}}\int_{-\infty}^{+\infty}e^{+\frac{i}{\hbar}ET}\ket{\vec{\chi}_T}dT.
\end{eqnarray}
The energy spectrum of Dirac equation is well-known and given by $E=\pm c\sqrt{(mc)^2+p^2}$, which can be obtained readily in the momentum space. Calculation of the time spectrum can also be done in the momentum-space, using the substituions $\hat{\textbf{p}}\rightarrow\textbf{p}$ and $\hat{\textbf{r}}\rightarrow i\hbar\frac{\partial}{\partial {\textbf{p}}}$. This leads to the matrix differential eigenvalue equation
\begin{eqnarray}\nonumber
i\hbar\begin{bmatrix}
             \frac{m}{6p^2}(2{\bf p}\cdot \frac{\partial}{\partial {\textbf{p}}}-3) & \frac{1}{3c}\vec{\sigma}\cdot \frac{\partial}{\partial {\textbf{p}}} \\
             \frac{1}{3c}\vec{\sigma}\cdot \frac{\partial}{\partial {\textbf{p}}} & -\frac{m}{6p^2}(2{\bf p}\cdot \frac{\partial}{\partial {\textbf{p}}}-3)
           \end{bmatrix}
\ket{\vec{\chi}_T({\bf p})}
\\
= T \ket{\vec{\chi}_T({\bf p})},\label{eq10}
\end{eqnarray}
\noindent
which can be investigated by numerical methods.
\begin{figure}[htp]
	\centering
	\includegraphics[width=3.25in]{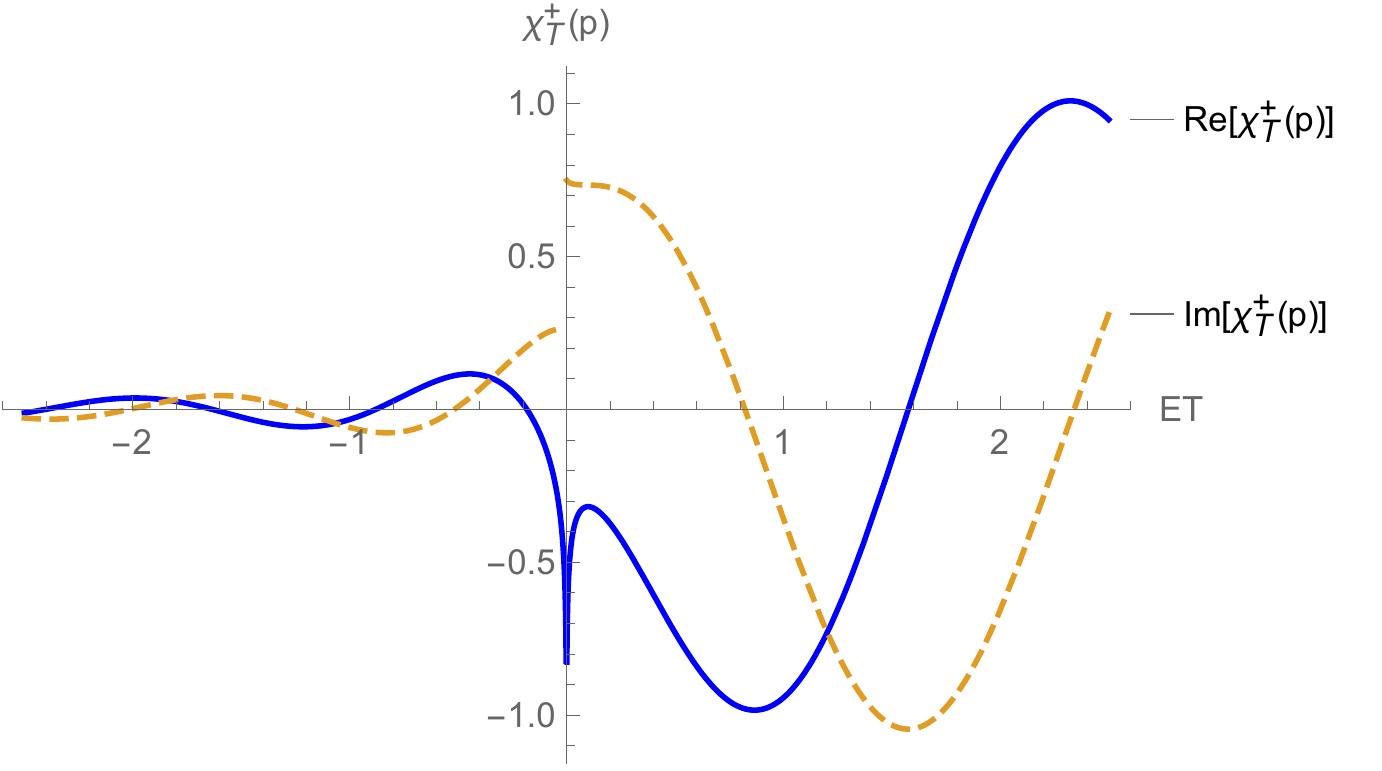}
	\caption{Real (solid) and imaginary (dashed) values of the time eigenfunction of particles versus \textcolor{black}{energy time product $ET$, which is in units of $\hbar$}.}\label{Fig1}
\end{figure}

\begin{figure}[htp]
	\centering
	\includegraphics[width=3.25in]{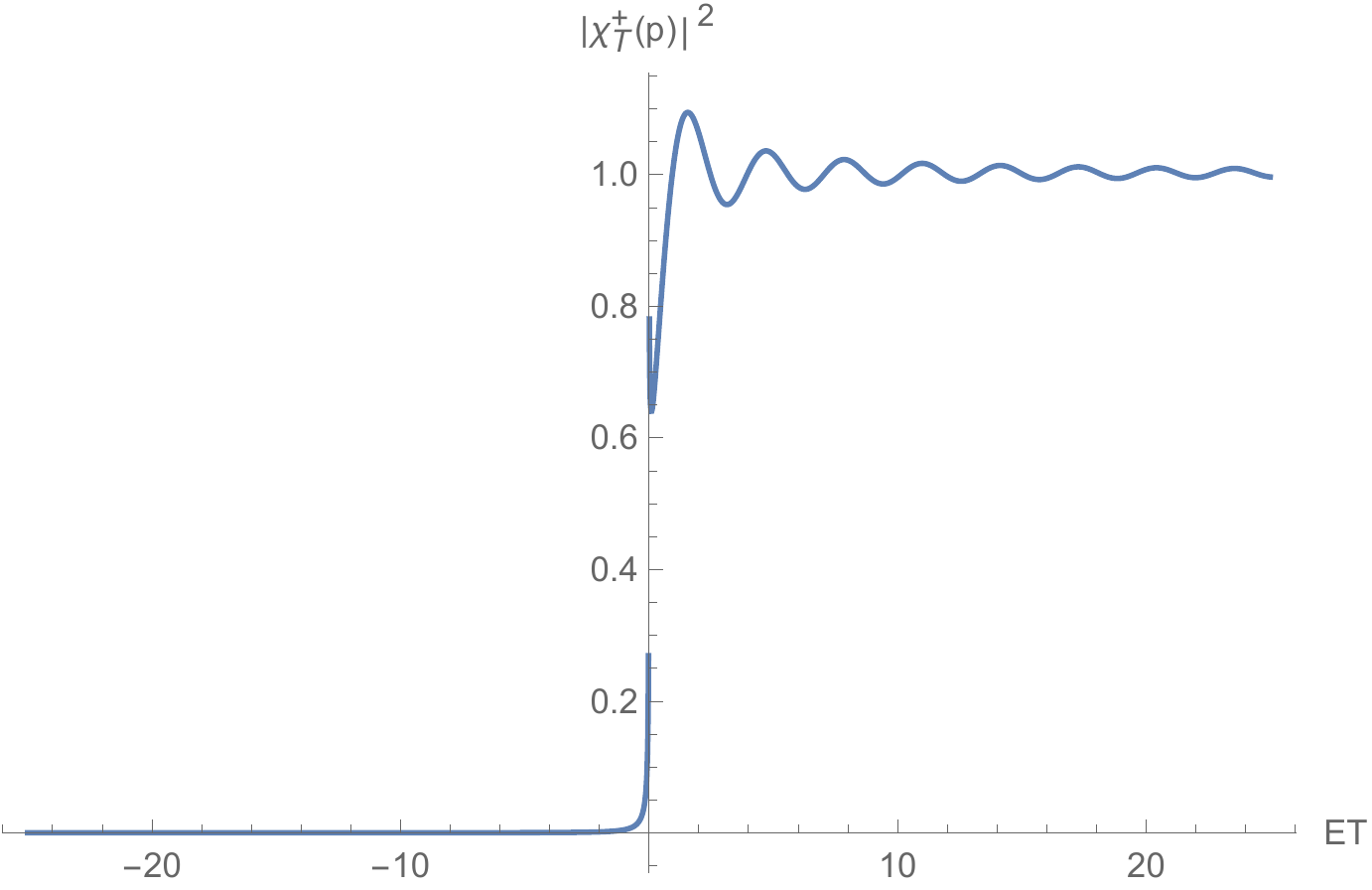}
	\caption{Squared absolute value of the time eigenfunction of particles versus \textcolor{black}{energy time product $ET$}. The probablity density function decays rapidly to zero for negative \textcolor{black}{$ET$, which is in units of $\hbar$}.}\label{Fig2}
\end{figure}

Quite remarkably, in the non-relativistic limit with $c\rightarrow\infty$, the time-operator is diagonalized and for particles reduces to the scalar form
\begin{equation}\label{eq11}
\mathbb{T}=\frac{m}{6\hat{p}^2}(3\hat{\bf p}\cdot\hat{\bf r}-\hat{\bf r}\cdot\hat{\bf p}),
\end{equation}
\noindent
which also satisfies the commutation $[\mathbb{T},\mathbb{K}]=i\hbar$ with the Newtonian kinetic energy operator $\mathbb{K}=\frac{1}{2m}p^2$. The time-eigenvalues here constitute a continuous spectrum over the entire real axis, while the time-eigenfunctions are found \textcolor{black}{following (\ref{eq8})} from the first-order nonlinear differential equation
\begin{equation}\label{eq12}
\textcolor{black}{
\frac{m}{6p^2}(3{\bf p}\cdot \frac{\partial}{\partial {\textbf{p}}}-1)\chi_T({\bf p})=-\frac{i}{\hbar}T\chi_T({\bf p})}.
\end{equation}
\noindent
The equation (\ref{eq12}) admits an exact solution
\begin{eqnarray}\label{eq13}
\textcolor{black}{
  \chi_T({\bf p})}&\textcolor{black}{=}&\textcolor{black}{\alpha\exp(-\frac{iTp^2}{\hbar m})[\ei(\frac{iTp^2}{\hbar m})+i\beta]}\\ \nonumber
&\textcolor{black}{=}&\textcolor{black}{\alpha\exp(-\frac{i2ET}{\hbar})[\ei(\frac{i2ET}{\hbar})+i\beta]},
\end{eqnarray}
\noindent
with $\ei(\cdot)$ being the Euler's exponential integral function and $\beta$ is a constant, determining the initial conditions at zero \textcolor{black}{time} and $\alpha$ is a normalization constant. \textcolor{black}{It is here furthermore noticed that for a non-relativistic massive particle in free space, the kinetic energy is simply $E=p^2/2\hbar m$.} 

\begin{figure}[htp]
	\centering
	\includegraphics[width=3.25in]{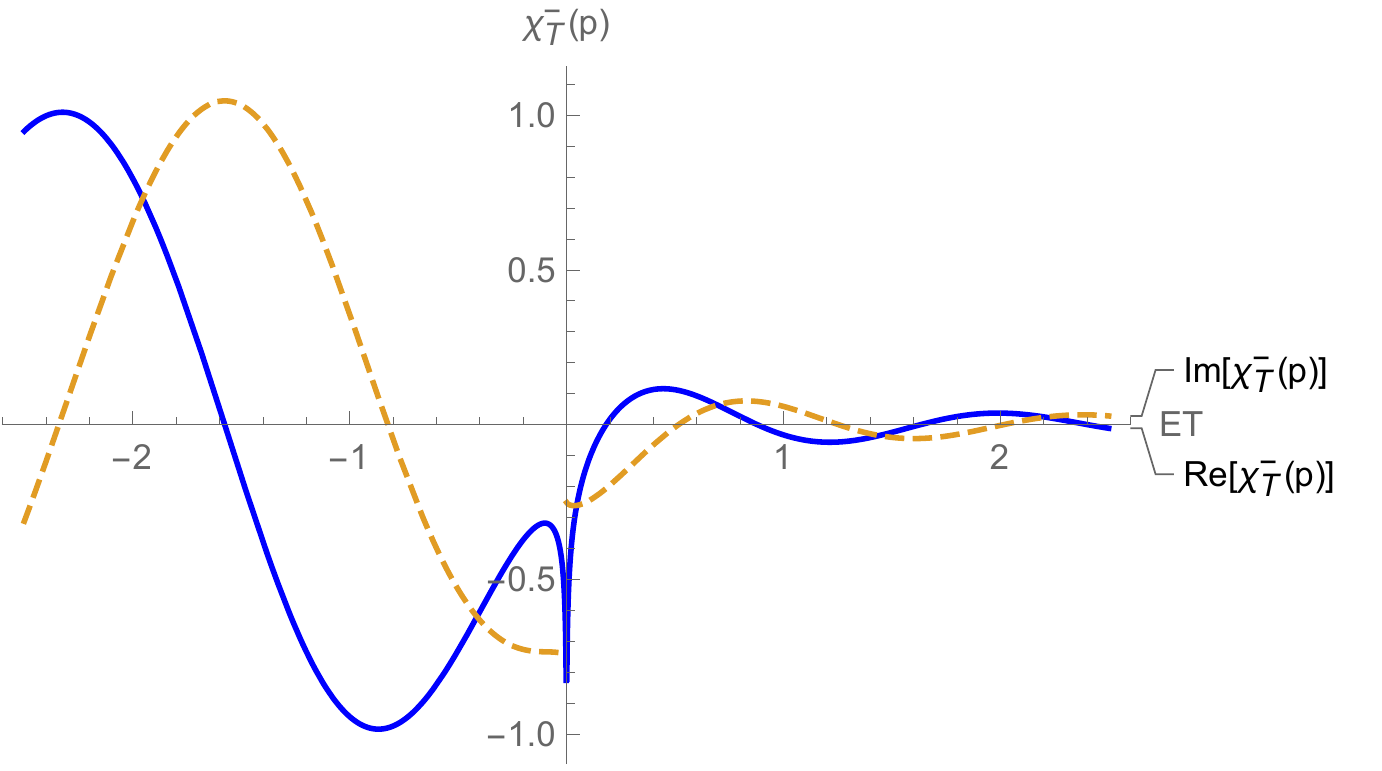}
	\caption{Real (solid) and imaginary (dashed) values of the time eigenfunction of antiparticles versus \textcolor{black}{energy time product $ET$}. Both components decays to zero for positive \textcolor{black}{$ET$, which is in units of $\hbar$}.}\label{Fig3}
\end{figure}

\begin{figure}[htp]
	\centering
	\includegraphics[width=3.25in]{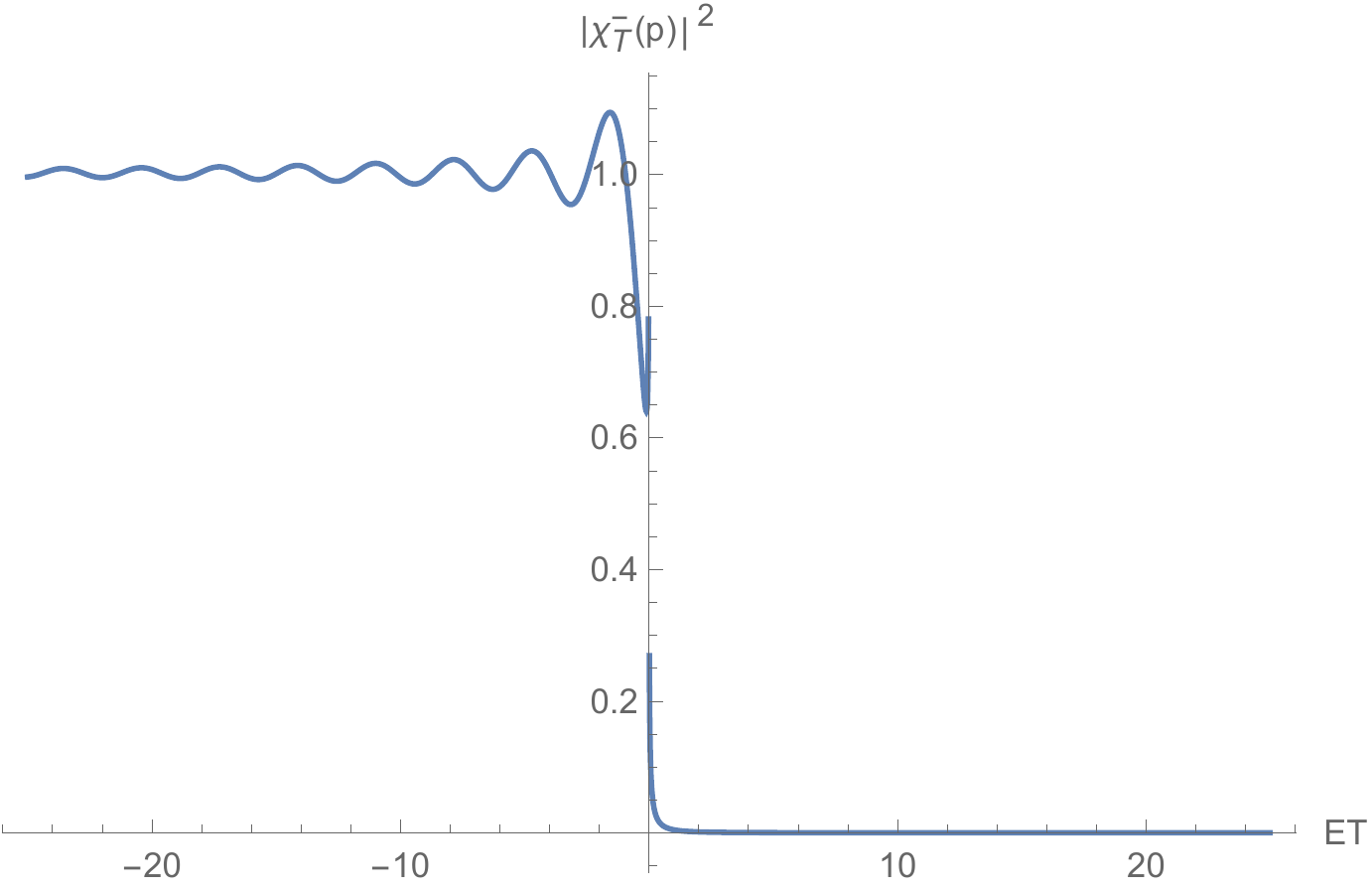}
	\caption{Squared absolute value of the time eigenfunction of antiparticles versus \textcolor{black}{energy time product $ET$}. The probablity density function decays rapidly to zero for positive \textcolor{black}{$ET$, which is in units of $\hbar$}.}\label{Fig4}
\end{figure}

Setting $\beta=\pm \pi$ differentiates the solutions corresponding to particles and anti-particles, occupying the positive or negative \textcolor{black}{energy time products}, as
\begin{eqnarray}\label{eq14}
\textcolor{black}{
  \chi_T^+({\bf p})=\alpha\exp(-\frac{iTp^2}{\hbar m})[\ei(\frac{iTp^2}{\hbar m})+i\pi]},\\ \nonumber
\textcolor{black}{\chi_T^-({\bf p})=\alpha\exp(-\frac{iTp^2}{\hbar m})[\ei(\frac{iTp^2}{\hbar m})-i\pi]}.
\end{eqnarray}
Here, we set $\beta=+\pi$ for particles and obtain the particle time eigenfunctions. \textcolor{black}{Figures \ref{Fig1} and \ref{Fig2}} illustrate the variations of particle time eigenfunctions versus \textcolor{black}{energy time product} in momentum space. Figures \ref{Fig3} and \ref{Fig4} are the same but for antiparticles. 

As shown in Fig. \ref{Fig2}, and since the \textcolor{black}{causal} particles \textcolor{black}{at positive times} cannot attain negative energies, the probablity density function decays rapidly to zero, while it quickly attains a nearly fixed value for positive \textcolor{black}{energy time product}. This reveals that particles can indeed have negative energies, too, but with vanishingly small probablity, while the occupation probablity for positive energies is also not accurately constant around zero energy. \textcolor{black}{Similarly, anticausal antiparticles at positive times cannot attain positive energies, and their probablity density function decays rapidly to zero. Also, antiparticles can indeed have positive energies, but with vanishingly small probablity. }

Therefore, with proper choice of the normalization constant $\alpha$, we may approximate the probablity density of time eigenfunctions in momentum space simply as \textcolor{black}{$|\chi_T^+({\bf p})|^2\approx u(+T)$} for particles, and \textcolor{black}{$|\chi_T^-({\bf p})|^2\approx u(-T)$} for antiparticles, with $u(\cdot)$ being the unit-step function. This, of course, not only quite reasonably meets the generally expected behavior of classical physics in the limit of $\hbar \rightarrow 0$, but also resolves the long-held assertion of Wolfgang Pauli's debate \cite{5} regarding the existence of a self-adjoint time operator, that the energy spectrum should be bounded from below. It is possible that findings of this paper could have immediate use in the theory of ultrarelativistic neutrino oscillations \cite{21,22,23} and other spin-$\frac{1}{2}$ particles \cite{24} as well. 

As a final remark, the time operator (\ref{eq11}) has been obtained from the non-relativistic limit of (\ref{eq10}), which removes any ambiguity with regard to the nonuniqueness of the time operator \cite{25}. Having both the Hamiltonian $\mathbb{K}$ and time operator $\mathbb{T}$ known, we may construct a new bosonic quasi-particle field \cite{26} out of the Harmonic-oscillator system \cite{6} as
\begin{eqnarray}
\label{eq15}
\mathbb{F}&=&mc^2\left(\hat{f}^\dagger\hat{f}+\frac{1}{2}\right),\\ \nonumber
\hat{f}&=&\frac{1}{\sqrt{2}}\left(\frac{1}{mc^2}\mathbb{K}-i\frac{mc^2}{\hbar}\mathbb{T}\right),\\ \nonumber
\hat{f}^\dagger&=&\frac{1}{\sqrt{2}}\left(\frac{1}{mc^2}\mathbb{K}+i\frac{mc^2}{\hbar}\mathbb{T}\right),
\end{eqnarray} 
\noindent
in which $\hat{f}$ and $\hat{f}^\dagger$ obviously satisfy $[\hat{f},\hat{f}^\dagger]=1$ because of $[\mathbb{T},\mathbb{K}]=i\hbar$, and are respectively the annihilation and creation operators of the bosonic quasi-particles, which we here refer to as \textit{energons}.

\begin{figure}[htp]
	\centering
	\includegraphics[width=3.25in]{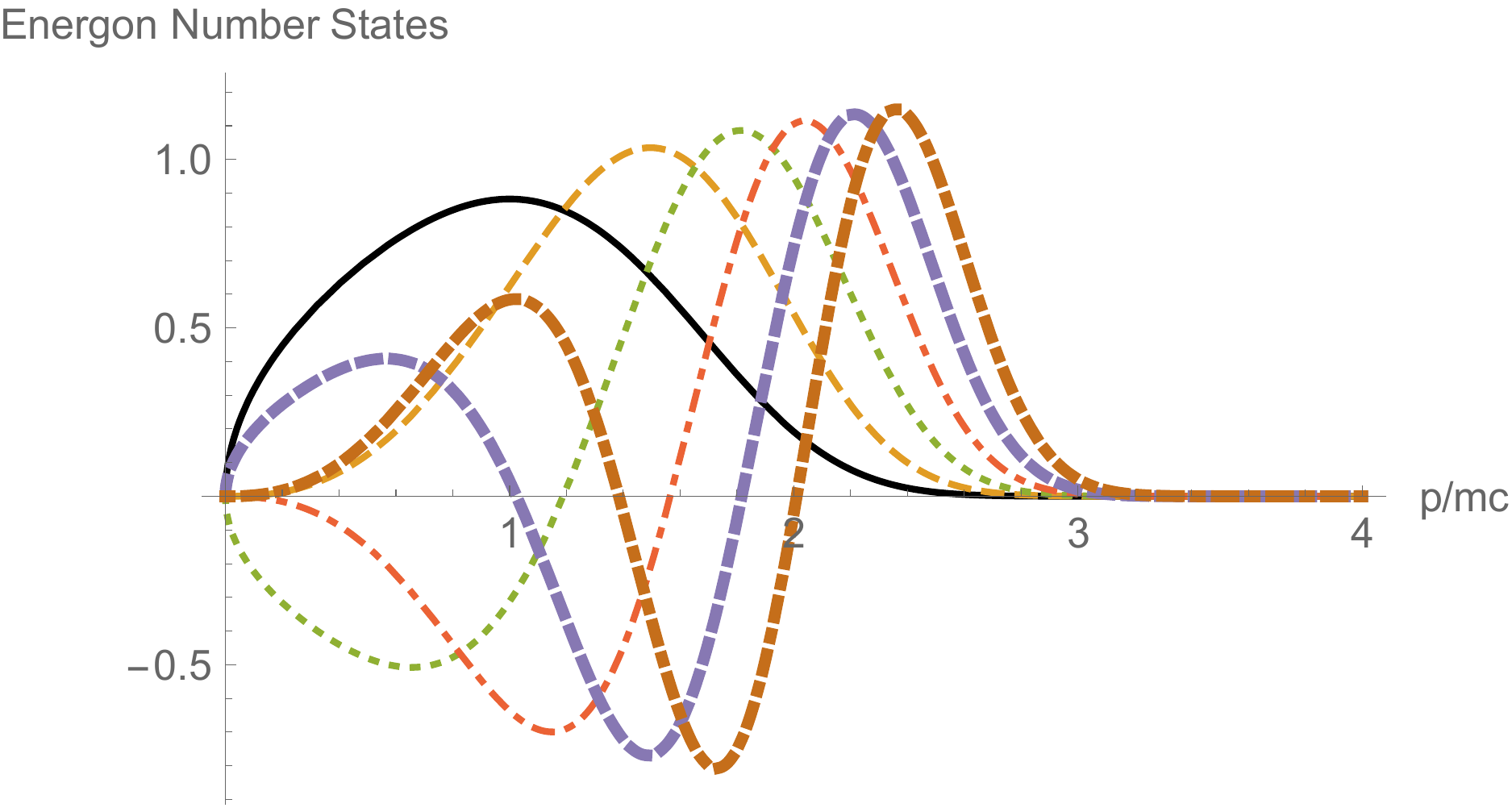}
	\caption{Energon number states: $\ket{0}$: black; $\ket{1}$: dashed; $\ket{2}$: dotted; $\ket{3}$: dot-dashed; $\ket{4}$: thick dashed; $\ket{5}$: thick dot-dashed.}\label{Fig5}
\end{figure}

In a strictly one-dimensional (1D) case, the time operator (\ref{eq11}) has to change a bit as $\mathbb{T}=\frac{m}{2\hat{p}^2}(3\hat{p}\hat{r}-\hat{r}\hat{p})$ and this allows to write down the differential equation for the zero energon $\hat{f}\ket{0}=\ket{\o}$ in momentum space $p\in[0,\infty)$ as
\begin{equation}
\label{eq16}
\frac{1}{\sqrt{2}}\left[\frac{1}{2m^2c^2}p^2+\frac{m^2c^2}{2p^2}(2p\frac{\partial}{\partial p}-1)\right]\zeta_0(p)=0,
\end{equation}   
\noindent
with $\zeta_0(p)=\braket{p}{0}$ being the momentum representation of the ground state with zero number of energons. The normalized ground-state solution is 
\begin{equation}
\label{eq17}
\zeta_0(p)=\frac{\sqrt{2p}}{\pi^\frac{1}{4}mc} \exp\left[-\frac{1}{8}\left(\frac{p}{mc}\right)^4\right].
\end{equation}
\noindent
By successive application of $\hat{f}^\dagger$ to the ground state $\ket{0}$, the next number states could be easily constructed using $\hat{f}^\dagger\ket{n}=\sqrt{n+1}\ket{n+1}$. Similarly, of course, we have the conjugate ladder identity as $\hat{f}\ket{n}=\sqrt{n}\ket{n-1}$. For instance, we may observe that
\begin{eqnarray}
\label{eq18}
\zeta_1(p)&=&\frac{p^2\sqrt{p}}{\pi^\frac{1}{4}(mc)^3} \exp\left[-\frac{1}{8}\left(\frac{p}{mc}\right)^4\right],\\ \nonumber
\zeta_2(p)&=&\frac{p^4-2(mc)^4}{2\pi^\frac{1}{4}(mc)^5}\sqrt{p} \exp\left[-\frac{1}{8}\left(\frac{p}{mc}\right)^4\right],\\ \nonumber
\zeta_3(p)&=&\frac{p^4-6(mc)^4}{4\sqrt{3}\pi^\frac{1}{4}(mc)^7}p^2\sqrt{2p} \exp\left[-\frac{1}{8}\left(\frac{p}{mc}\right)^4\right],
\end{eqnarray}
\noindent
and so on.

The question of whether energons are plain mathematical artifacts, or could possibly have a physical meaning, needs further investigation in detail, which remains as the subject of a future study.

\end{document}